\documentclass[12pt]{article}
\usepackage{amsfonts}
\usepackage{bm}
\usepackage{graphicx}
\usepackage{multicol}
\usepackage{color}
\begin{document}

\title{Minimal momentum estimation in noncommutative phase space of canonical type with preserved rotational and time reversal symmetries}
\maketitle

\centerline {Kh. P. Gnatenko \footnote{E-Mail address: khrystyna.gnatenko@gmail.com}}
\medskip

\centerline {\small  \it  Ivan Franko National University of Lviv,}
\centerline {\small \it Department for Theoretical Physics,12 Drahomanov St., Lviv, 79005, Ukraine}

\begin{abstract}
 Noncommutative algebra which is rotationally invariant, time reversal invariant and equivalent to noncommutative algebra of canonical type is considered. Perihelion shift of orbit of a particle in Coulomb potential in the rotationally-invariant noncommutative phase space is found up to the second order in the parameters of noncommutativity. Applying the result to the case of Mercury planet and using observable results for precession of its orbit we find upper bounds on the parameters of noncommutativity in the rotationally-invariant noncommutative phase space. The obtained upper bound for the parameter of momentum noncommutativity is at least ten orders less than the upper bounds estimated on the basis of studies of the hydrogen atom in noncommutative phase space.  As a result we obtain stringent restriction on for the minimal momentum in noncommutative phase space with preserved rotational and time reversal symmetries.

\end{abstract}

\section{Introduction}

In account of development of String theory and Quantum gravity (see, for example, \cite{Witten,Doplicher}), studies of modifications of commutation relations leading to the minimal length have received much attention. Idea that commutator of coordinates may not be equal to zero was proposed by Heisenberg. The first paper on the subject was written by Snyder  \cite{Snyder}.
Noncommutative phase space of canonical type is characterized by the following commutation relations
  \begin{eqnarray}
[X_{i},X_{j}]=i\hbar\theta_{ij},\label{form101}\\{}
[X_{i},P_{j}]=i\hbar(\delta_{ij}+\gamma_{ij}),\label{form1001}\\{}
[P_{i},P_{j}]=i\hbar\eta_{ij},\label{form10001}{}
\end{eqnarray}
where $\theta_{ij}$, $\eta_{ij}$, $\gamma_{ij}$ are elements of constant  matrixes.

It is known that noncommutativity of canonical type (\ref{form101})-(\ref{form10001}) causes breaking of rotational and time reversal symmetries  \cite{Chaichian,Balachandran,Geloun,Scholtz,GnatenkoPRA}. To recover the rotational symmetry different generalizations of commutation relations were considered as a result different types of noncommutative algebras were proposed \cite{Moreno,Galikova,Amorim,GnatenkoPLA14}.
Studies of rotationally-invariant noncommutative algebra with position-dependent noncommutativity  (see, for instance, \cite{Lukierski,Lukierski2009,BorowiecEPL,Borowiec,Borowiec1,Kupriyanov2009,Kupriyanov}), rotationally-invariant noncommutative algebra with involving spin degrees of freedom (see, for instance, \cite{Falomir09,Ferrari13,Deriglazov}) have received much attention.   Algebra which is rotationally and time reversal invariant and besides equivalent to noncommutative algebra of canonical type was constructed in \cite{GnatenkoPRA}.

Influence of space quantization on the perihelion shift of a particle in Coulomb potential was examined in  noncommutative space of canonical type \cite{Romero,Mirza}, noncommutative phase space of canonical type \cite{Djemai2,Gnatenko_arxiv}, deformed space with minimal length \cite{Benczik02,Silagadze}, Snyder space \cite{Ivetic}.  The results were  used for estimation of the values of parameters of the corresponding algebras and for setting upper bound on the minimal length.  The upper bounds for the minimal length presented in \cite{Benczik02,Ivetic,Romero,Mirza} are extremely small (they are many orders less than the Planck length). These bounds were reexamined to more relevant one taking into consideration that commutation relations for coordinates and momenta of the center-of-mass of macroscopic body are not the same as commutation relations for coordinates and momenta of a particle \cite{Tk2,GnatenkoPLA13,GnatenkoEPL19}. It is worth noting that, taking into account features of noncommutative algebra for coordinates and momenta of the center-of-mass, in the paper \cite{Gnatenko_arxiv} stringent upper bound on the momentum scale in noncommutative phase space of canonical type (\ref{form101})-(\ref{form10001}) was obtained.

In the present paper we examine influence of noncommutativity of coordinates and noncommutativity of momenta on the perihelion shift of a particle in Coulomb potential in the frame of rotationally and time reversal invariant noncommutative algebra of canonical type proposed in \cite{GnatenkoPRA}. The results are generalized to the case of macroscopic body. Taking into consideration features of noncommutative algebra for coordinates and momenta of the center-of-mass of a body in rotationally-invariant noncommutative phase space, we find perihelion shift of its orbit caused by noncommutativity of coordinates and noncommutativity of momenta. Using data for precession of Mercury's perihelion from ranging to the MESSENGER spacecraft \cite{Park} we estimate upper bound on the parameters of noncommutativity in the space. The obtained result for the parameter of momentum noncommutativity is quite strong and improves result obtained on the basis of studies of the hydrogen atom in rotationally-invariant noncommutative phase  space \cite{GnatenkoIJMPA17}.

The paper is organized as follows. In Section 2 noncommutative algebra which is rotationally-invariant and invariant under the time reversal is presented and features of description of composite system motion in the frame of the algebra are discussed. Section 3 is devoted to studies of influence of noncommutativity of coordinates and noncommutativity of momenta on the perihelion shift of a particle in the noncommutative phase space. The result is generalized to the case of macroscopic body. The upper bounds for the parameter of coordinate noncommutativity and parameter of momentum noncommutativity are estimated in Section 4. Conclusions are presented in Section 5.

\section{Noncommutative algebra of canonical type with preserved rotational and time reversal symmetries}

Noncommutative algebra which is rotationally and time reversal invariant reads
\begin{eqnarray}
[X_{i},X_{j}]=i\hbar\theta_{ij}=ic_{\theta}\sum_k\varepsilon_{ijk}{p}^a_{k},\label{for101}\\{}
[X_{i},P_{j}]=i\hbar(\delta_{ij}+\gamma_{ij})=i\hbar\left(\delta_{ij}+\frac{c_{\theta}c_{\eta}}{4\hbar^2}({\bf {p}}^a\cdot{\bf {p}}^b)\delta_{ij}-\frac{c_{\theta}c_{\eta}}{4\hbar^2}p^{a}_j{p}^{b}_i\right),\label{for1001}\\{}
[P_{i},P_{j}]=i\hbar\eta_{ij}=i{c_{\eta}}\sum_k\varepsilon_{ijk}{p}^{b}_{k}.{}\label{for10001}
\end{eqnarray}
Algebra (\ref{for101})-(\ref{for10001}) is constructed in \cite{GnatenkoPRA} on the basis of idea of generalization of parameters of noncommutativity to tensors defined as
 \begin{eqnarray}
\theta_{ij}=\frac{c_{\theta}}{\hbar}\sum_k\varepsilon_{ijk}{p}^a_{k},\label{t1}\\
 \eta_{ij}=\frac{c_{\eta}}{\hbar}\sum_k\varepsilon_{ijk}{p}^b_{k},\label{t2}\\
 \gamma_{ij}=\sum_k \frac{\theta_{ik}\eta_{jk}}{4}=\frac{c_{\theta}c_{\eta}}{4\hbar^2}({\bf {p}}^a\cdot{\bf {p}}^b)\delta_{ij}-\frac{c_{\theta}c_{\eta}}{4\hbar^2}p^{a}_j{p}^{b}_i.\label{g1}
 \end{eqnarray}
 In (\ref{for101})-(\ref{g1}) $c_{\theta}$, $c_{\eta}$ are constants (in the classical limit $\lim_{\hbar\rightarrow0}c_{\theta}/\hbar=const$, $\lim_{\hbar\rightarrow0}c_{\eta}/\hbar=const$), ${p}^{a}_i$, ${p}^{b}_i$ are additional momenta.
For preserving of the rotational symmetry additional coordinates and momenta ${a}_i$, ${b}_i$  ${p}^a_i$, ${p}^b_i$ are assumed to be governed by rotationally-symmetric systems. For simplicity these systems are considered to be harmonic oscillators
\begin{eqnarray}
H^a_{osc}=\frac{({\bf p}^{a})^{2}}{2m_{osc}}+\frac{m_{osc}\omega^2_{osc}{\bf a}^{2}}{2}, \ \ H^b_{osc}=\frac{({\bf p}^{b})^{2}}{2m_{osc}}+\frac{m_{osc}\omega^2_{osc}{\bf b}^{2}}{2},\label{oscb}
\end{eqnarray}
with very large frequency $\omega_{osc}$. Therefore because of large distance between the energy levels the oscillators put into the ground states remain in the states.  The length of the oscillator is supposed to be equal to the Planck length $\sqrt{\hbar}/\sqrt{{m_{osc}\omega_{osc}}}=l_{P}$ \cite{GnatenkoPRA}.
 Definition for parameters $\gamma_{ij}$  (\ref{g1}) follows from the symmetric representation of noncommutative coordinates and noncommutative momenta (see, for instance, \cite{Bertolamih,Bertolami06,Djemai1,Djemai2}).

Commutation relations for additional coordinates ${a}_i$, ${b}_i$,  and additional momenta ${p}^a_i$, ${p}^b_i$ read
$[{a}_{i},{a}_{j}]=[{b}_{i},{b}_{j}]=[{a}_{i},{b}_{j}]=0$, $[{p}^{a}_{i},{p}^{a}_{j}]=[{p}^{b}_{i},{p}^{b}_{j}]=[{p}^{a}_{i},{p}^{b}_{j}]=0$,
$[{a}_{i},{p}^{a}_{j}]=[{b}_{i},{p}^{b}_{j}]=i\hbar\delta_{ij},$
$[{a}_{i},{p}^{b}_{j}]=[{b}_{i},{p}^{a}_{j}]=0,$
$[{a}_{i},X_{j}]=[{a}_{i},P_{j}]=[{p}^{b}_{i},X_{j}]=[{p}^{b}_{i},P_{j}]=0$. From the last equality follows that as in the case of canonical version of noncommutative algebra  (\ref{form101})-(\ref{form10001}) the tensors of noncommutativity commute with coordinates and momenta $[\theta_{ij},X_{k}]=[\theta_{ij},P_{k}]=[\eta_{ij},X_{k}]=[\eta_{ij},P_{k}]=0,$ and
$[\gamma_{ij},X_{k}]=[\gamma_{ij},P_{k}]=0$. In this sense the proposed algebra (\ref{for101})-(\ref{for10001}) is equivalent to noncommutative algebra of canonical type (\ref{form101})-(\ref{form10001})\cite{GnatenkoPRA}.

 The coordinates and momenta satisfying  (\ref{for101})-(\ref{for10001}) can be represented as
 \begin{eqnarray}
X_{i}=x_{i}+\frac{1}{2}[{\bm \theta}\times{\bf p}]_i, \ \
P_{i}=p_{i}-\frac{1}{2}[{\bm \eta}\times {\bf x}]_i,\label{repp0}
\end{eqnarray}
where the components of vectors ${\bm \theta}$, ${\bm \eta}$ read
\begin{eqnarray}
\theta_i=\sum_{jk}\frac{\varepsilon_{ijk}\theta_{jk}}{2}=\frac{c_{\theta}{p}_i^a}{\hbar}, \ \  \eta_i=\sum_{jk}\frac{\varepsilon_{ijk}\eta_{jk}}{2}=\frac{c_{\eta} p_i^b}{\hbar}.
\end{eqnarray}
Coordinates and momenta $x_i$, $p_i$ satisfy $[x_i,x_j]=[p_i,p_j]=0, [x_i,p_j]=i\hbar\delta_{ij}$. Therefore, the Jacobi identity holds for all possible triplets of operators.

The relations of algebra  (\ref{for101})-(\ref{for10001}) remain the same after rotation. One has
\begin{eqnarray}
[X^{\prime}_{i},X^{\prime}_{j}]=ic_{\theta}\sum_k\varepsilon_{ijk}{p}^{a\prime}_{k},\label{fo101}\\{}
[X^{\prime}_{i},P^{\prime}_{j}]=i\hbar\left(\delta_{ij}+\frac{c_{\theta}c_{\eta}}{4\hbar^2}({\bf {p}}^{a\prime}\cdot{\bf {p}}^{b\prime})\delta_{ij}-\frac{c_{\theta}c_{\eta}}{4\hbar^2}p^{a\prime}_j{p}^{b\prime}_i\right),\label{fo1001}\\{}
[P^{\prime}_{i},P^{\prime}_{j}]=i{c_{\eta}}\sum_k\varepsilon_{ijk}{p}^{b\prime}_{k},{}\label{fo10001}
\end{eqnarray}
where  $X_{i}^{\prime}=U(\varphi)X_{i}U^{+}(\varphi)$, $P_{i}^{\prime}=U(\varphi)P_{i}U^{+}(\varphi)$ $p_{i}^{a\prime}=U(\varphi)p^a_{i}U^{+}(\varphi)$,  $p^{b\prime}_{i}=U(\varphi)p^b_{i}U^{+}(\varphi)$ and the operator of rotation is defined as $U(\varphi)=\exp(i\varphi({\bf n}\cdot{\bf L^t})/\hbar)$ with ${\bf L^t}=[{\bf x}\times{\bf p}]+[{\bf{a}}\times{\bf {p}}^{a}]+[{\bf{b}}\times{\bf { p}}^{b}]$, $U^{+}(\varphi)=\exp(-i\varphi({\bf n}\cdot{\bf L^t})/\hbar)$  \cite{GnatenkoIJMPA17}.
So, the algebra (\ref{for101})-(\ref{for10001}) is rotationally-invariant.

Upon time reversal coordinates and momenta transform as $X_{i}\rightarrow X_{i}$, $P_{i}\rightarrow -P_{i}$, $p^a_i\rightarrow-p^a_i$, $p^b_i\rightarrow-p^b_i$, therefore $\theta_{ij}\rightarrow -\theta_{ij}$, $\eta_{ij}\rightarrow -\eta_{ij}$. So, algebra  (\ref{for101})-(\ref{for10001})  remains the same after the time reversal \cite{GnatenkoPRA}.

For description of motion of a composite system  (macroscopic body) in the noncommutative phase space relations (\ref{for101})-(\ref{for10001})  can be generalized as
 \begin{eqnarray}
[X^{(n)}_{i},X^{(m)}_{j}]=i\hbar\delta_{mn}\theta^{(n)}_{ij},\label{ffor101}\\{}
[X^{(n)}_{i},P^{(m)}_{j}]=i\hbar\delta_{mn}\left(\delta_{ij}+\sum_k\frac{\theta^{(n)}_{ik}\eta^{(m)}_{jk}}{4}\right),\label{for1001}\\{}
[P^{(n)}_{i},P^{(m)}_{j}]=i\hbar\delta_{mn}\eta^{(n)}_{ij},\label{ffor10001}
 \end{eqnarray}
where indexes $m,n$ label the particles and $\theta^{(n)}_{ij}$, $\eta^{(n)}_{ij}$ are tensors of noncommutativity corresponding to the particle labeled by index $n$.

In the frame of rotationally-invariant noncommutative algebra (\ref{for101})-(\ref{for10001}) with tensors of noncommutativity defined as $\theta_{ij}={l_0}\sum_k\varepsilon_{ijk}{a}_{k}/\hbar$, $\eta_{ij}={p_0}\sum_k\varepsilon_{ijk}{p}^b_{k}/\hbar$ (here $l_0$, $p_0$ are constants and $a_k$, $p^b_k$ are additional coordinates and momenta governed by harmonic oscillators (\ref{oscb})) the problem of description of a composite system motion has been studied in our previous paper \cite{GnatenkoIJMPA18}.
 In the paper \cite{GnatenkoIJMPA18} we concluded that
if tensor of coordinate noncommutativity  is proportional to  mass and tensor of momentum noncommutativity is proportional inversely to mass,  commutation relations for coordinates and momenta of the center-of-mass reproduce noncommutative algebra for coordinates and momenta of individual particles, noncommutative coordinates can be considered as kinematic variables and noncommutative momenta are proportional to mass. Besides in  \cite{GnatenkoEPL18} we show that these relations of tensors of noncommutativity with mass open possibility to recover the weak equivalence principle  in the rotationally-invariant noncommutative phase space.
 Similar conclusions can be done in the frame of algebra (\ref{for101})-(\ref{for10001}).

We consider tensors of noncommutativity to be dependent on mass and to be defined as
  \begin{eqnarray}
\theta^{(n)}_{ij}=\frac{c_{\theta}^{(n)}}{\hbar}\sum_k\varepsilon_{ijk}{p}^{a}_{k}, \ \  \eta^{(n)}_{ij}=\frac{c_{\eta}^{(n)}}{\hbar}\sum_k\varepsilon_{ijk}{p}^b_{k}, \label{tc2}
 \end{eqnarray}
 where the constants $ c^{(n)}_{\theta}$, $ c^{(n)}_{\eta}$ satisfy the following relations
  \begin{eqnarray}
 c^{(n)}_{\theta}m_n=\tilde{\gamma},\label{condt}\\
\frac{c^{(n)}_{\eta}}{m_n}=\tilde{\alpha},\label{conde}
 \end{eqnarray}
here  $\tilde{\gamma}$, $\tilde{\alpha}$ are constants which do to depend on mass.
Additional coordinates $a_i$, $b_i$ and additional momenta $p^a_i$, $p^b_i$ are responsible for the noncommutativity of the phase space. Particles corresponds to
the same noncommutative phase space. Therefore, additional momenta  $p^a_i$, $p^b_i$ in  (\ref{tc2}) are considered to be the same for different particles. According to  (\ref{condt}), (\ref{conde}), effect of noncommutativity on the particles with different masses is different.

Note that if conditions (\ref{condt}), (\ref{conde}) hold,
relations for coordinates and momenta of the center-of-mass ${\bf X}^c=\sum_{n}\mu_{n}{\bf X}^{(n)}$, ${\bf P}^c=\sum_{n}{\bf P}^{(n)}$ (here $\mu_n=m_n/M$, $M=\sum_nm_n$)  reproduce relations of noncommutative algebra (\ref{ffor101})-(\ref{ffor10001})
 \begin{eqnarray}
[X^c_i,X^c_j]=i\hbar{\theta}^c_{ij}, \ \ [P^c_i,P^c_j]=i\hbar{\eta}^c_{ij},\label{xc}\\{}
[{X}^c_i,{P}^c_j]=i\hbar(\delta_{ij}+\sum_k\frac{\theta^c_{ik}\eta^c_{jk}}{4}).\label{pc}{}
\end{eqnarray}
Here effective tensors of noncommutativity ${\theta}^c_{ij}$, ${\eta}^c_{ij}$ read
 \begin{eqnarray}
 {\theta}^c_{ij}=\sum_n \mu^2_n\theta_{ij}^{(n)}=\frac{\tilde{\gamma}}{\hbar M}\sum_k\varepsilon_{ijk}p^{a}_{k},\label{effc}\\
{\eta}^c_{ij}=\sum_{n}\eta^{(n)}_{ij}=\frac{\tilde{\alpha}M}{\hbar}\sum_k\varepsilon_{ijk}{p}^b_{k}.\label{eff2c}
 \end{eqnarray}

Representation for coordinates and momenta satisfying  (\ref{ffor101})-(\ref{ffor10001}) reads
 \begin{eqnarray}
X^{(n)}_{i}=x^{(n)}_{i}+\frac{1}{2}[{\bm \theta}^{(n)}\times{\bf p}^{(n)}]_i, \ \
P^{(n)}_{i}=p^{(n)}_{i}-\frac{1}{2}[{\bm \eta}^{(n)}\times {\bf x}^{(n)}]_i,\label{rep}
\end{eqnarray}
where coordinates $x^{(n)}_{i}$ and momenta $p^{(n)}_{i}$ satisfy the ordinary commutation relations
\begin{eqnarray}
[x^{(n)}_{i},x^{(m)}_{j}]=[p^{(n)}_{i},p^{(m)}_{j}]=0,\ \ [x^{(n)}_{i},p^{(m)}_{j}]=i\hbar\delta_{ij}\delta_{mn}.\label{ord}
\end{eqnarray}
Note, that according to (\ref{rep}) coordinates depend on momenta and therefore they depend on mass.
If conditions (\ref{condt}), (\ref{conde})  hold,  the noncommutative coordinates do not depend on mass and noncommutative momenta are proportional to mass as it should be.  Taking into account  (\ref{condt}), (\ref{conde}), (\ref{rep}),   we can write
\begin{eqnarray}
X^{(n)}_{i}=x^{(n)}_{i}+\frac{\tilde{\gamma}}{2\hbar} [{\bf p}^{a}\times\frac{{\bf p}^{(n)}}{m_n}]_i,\ \
P^{(n)}_{i}=p^{(n)}_{i}-\frac{m_n\tilde{\alpha}}{\hbar}[{\bf p}^{b}\times{\bf x}^{(n)}]_i.\label{repn}
\end{eqnarray}

In the next sections these conclusions are used for studies of influence of noncommutativity on the  Mercury's motion  and for estimation of the upper bounds on the parameters of noncommutativity.

\section{Effect of noncommutativity of coordinates and noncommutativity of momenta on the perihelion shift}\label{rozd3}

Let us consider a particle of mass $m$ in the gravitational filed $-k/X$ ($X=|{\bf X}|$, $k$ is a constant) in noncommutative phase space with preserved rotational and time-reversal symmetries (\ref{for101})-(\ref{for10001}) and find the perihelion shift of its orbit caused by the noncommutativity of coordinates and noncommutativity of momenta. Because of involving of additional momenta for construction of the tensors of noncommutativity (\ref{t1}), (\ref{t2})  we have to study the Hamiltonian which is the sum of Hamiltonian corresponding to the particle and Hamiltonians corresponding to the harmonic oscillators
\begin{eqnarray}
H=H_p+H^a_{osc}+H^b_{osc},\\
H_p=\frac{P^2}{2m}-\frac{mk}{X},
 \end{eqnarray}
here $X_i$, $P_i$ satisfy relations (\ref{for101})-(\ref{for10001}), terms $H^a_{osc}$, $H^b_{osc}$ are given by  (\ref{oscb}).
It is convenient to use  representation  (\ref{repp0}) and rewrite the total Hamiltonian as follows
\begin{eqnarray}
H=H_0+\Delta H,\\
H_0=\langle H_p\rangle_{ab}+H^a_{osc}+H^b_{osc}=\frac{{ p}^2}{2m}-\frac{mk}{x}+\frac{\langle\eta^2\rangle  x^2}{12m}-\frac{\langle\theta^2\rangle mk L^2}{8 x^5}+\nonumber\\+\frac{\langle\theta^2\rangle mk}{24}\left(\frac{1}{x^2} p^2\frac{1}{x}+\frac{1}{x}p^2\frac{1}{x^2}+\frac{\hbar^2}{x^5}\right)+H^a_{osc}+H^b_{osc},\label{he}\\
\Delta H= H-H_0=H_p-\langle H_p\rangle_{ab}=-\frac{({\bm \eta}\cdot{\bf L})}{2m}+\frac{[{\bm \eta}\times{\bf x}]^2}{8m}-\frac{m k}{2x^3}({\bm \theta}\cdot{\bf L})-\nonumber\\-\frac{\langle\eta^2\rangle {x}^2}{12m}+\frac{m k L^2\langle\theta^2\rangle}{8 x^5}+\frac{m k}{16}\left(\frac{1}{x^2}[{\bm \theta}\times{\bf p}]^2\frac{1}{x}+\frac{1}{x}[{\bm \theta}\times{\bf p}]^2\frac{1}{x^2}+\frac{\hbar^2}{x^7}[{{\bm \theta}}\times{\bf x}]^2\right)-\nonumber\\-\frac{3mk}{8x^5}({\bm \theta}\cdot{\bf L})^2-\frac{mk \langle\theta^2\rangle}{24}\left(\frac{1}{x^2} p^2\frac{1}{x}+\frac{1}{x}p^2\frac{1}{x^2}+\frac{\hbar^2}{x^5}\right),\label{dh}
\end{eqnarray}
here ${\bf L}=[{\bf x}\times{\bf p}]$, $x=|{\bf x}|$, $\langle...\rangle_{ab}$ denotes averaging over the eigenfunctions $\psi^{a}_{0,0,0}$, $\psi^{b}_{0,0,0}$ of harmonic oscillators $H_{osc}^a$, $H_{osc}^b$  in the ground states. Also in (\ref{he}), (\ref{dh}) we use notations
\begin{eqnarray}
\langle\theta^2\rangle=\sum_i\langle\theta^2_i\rangle=\sum_i\frac{c_{\theta}^2}{\hbar^2}\langle\psi^{a}_{0,0,0}|({p}^{a}_i)^2|\psi^{a}_{0,0,0}\rangle=\frac{3c_{\theta}^2}{2l_P^2},\label{thetar2}\\
\langle\eta^2\rangle=\sum_i\langle\eta^2_i\rangle=\sum_i\frac{c_{\eta}^2}{\hbar^2}\langle\psi^{b}_{0,0,0}| ({p}^{b}_i)^2|\psi^{b}_{0,0,0}\rangle=\frac{3c_{\eta}^2}{2l_P^2}.\label{etar2}
\end{eqnarray}
Expressions (\ref{he}), (\ref{dh}) are written up to the second order in the parameters of noncommutativity. The details of calculation of expansion over the parameters of noncommutativity and averaging over the eigenfunctions of harmonic oscillators needed to write (\ref{he}), (\ref{dh}) can be found in our previous papers \cite{GnatenkoPLA14,GnatenkoEPL18}.

Up to the second order in the $\Delta H$ one can study Hamiltonian $H_0$. Because up to the second order in the perturbation theory corrections to the spectrum of $H$ caused by $\Delta H$ vanish (see \cite{GnatenkoIJMPA18}). So, taking into account expression for $\Delta H$ (\ref{dh}) one has that up to the second order in the parameters of noncommutativity one can consider Hamiltonian $H_0$ (\ref{he}). Note also that  $H^a_{osc}$ and $H^b_{osc}$ commute with $\langle H_p\rangle_{ab}$.
So, up to the second order in the parameters of noncommutativity  in order to examine the classical motion of a particle in the Coulomb potential in noncommutative phase space (\ref{for101})-(\ref{for10001}) one can consider the following Hamiltonian
\begin{eqnarray}
\langle H_p\rangle_{ab}=\frac{{ p}^2}{2m}-\frac{mk}{x}+\frac{\langle\eta^2\rangle  x^2}{12m}-\frac{\langle\theta^2\rangle m k L^2}{8 x^5}+\frac{\langle\theta^2\rangle m k p^2}{12x^3}.\label{hec}
\end{eqnarray}
Because of terms in (\ref{hec}) caused by noncommutativity of coordinates and noncommutativity of momenta the orbit of the particle precesses.  Let us find the precession rate of perihelion of the orbit caused by the noncommutativity. For this purpose we consider the Hamilton vector defined as
\begin{eqnarray}
{\bf u}=\frac{{\bf p}}{m}-\frac{mk[{\bf L}\times{\bf x}]}{x L^2},
\end{eqnarray}
and calculate its precession rate
\begin{eqnarray}
{\bm \Omega}=\frac{[{\bf u}\times\dot{{\bf u}}]}{u^2}.
\end{eqnarray}
We have
\begin{eqnarray}
\left\{{\bf{u}},\frac{{ p}^2}{2m}-\frac{mk}{x}\right\}=0,
\end{eqnarray}
and for $ \dot{\bf{u}}$ we obtain
\begin{eqnarray}
 \dot{\bf{u}}=\left\{{\bf{u}},\frac{\langle\eta^2\rangle  x^2}{12m}-\frac{\langle\theta^2\rangle m k L^2}{8 x^5}+\frac{\langle\theta^2\rangle m k p^2}{12x^3}\right\}=\nonumber\\=-\frac{\langle\eta^2\rangle {\bf x}}{6m^2}-\frac{k\langle\theta^2\rangle}{4}\left(\frac{({\bf x}\cdot{\bf p}){\bf p}}{x^5}-\frac{2p^2{\bf x}}{x^5}+\frac{5L^2{\bf x}}{2x^7}\right)+\nonumber\\+\frac{m^2k^2\langle\theta^2\rangle[{\bf L}\times{\bf p}]}{12L^2{x}^4}-\frac{m^2k^2\langle\theta^2\rangle({\bf x\cdot{\bf p}})[{\bf L}\times{\bf x}]}{12L^2x^6}.\label{du}
\end{eqnarray}
Taking into account that in the ordinary space ($\theta_{ij}=\eta_{ij}=0$) $u^2=m^2 k^2e^2/L^2$ with $e$ being the eccentricity of the orbit, we find
 \begin{eqnarray}
{\Omega}=\langle\theta^2\rangle\left(\frac{5 L^4}{8km^3x^7e^2}-\frac{p^2L^2}{2m^3x^5k e^2}+\frac{p^2}{4me^2x^4}-\frac{7L^2}{24mx^6e^2}-\frac{m k}{12x^5e^2}\right){ L}+\nonumber\\+
\langle\eta^2\rangle\left(\frac{L^2}{6m^5k^2e^2}-\frac{x}{6m^3k e^2}\right){L}.
\end{eqnarray}
So, up to the second order in the parameters of noncommutativity the perihelion shift per revolution reads
\begin{eqnarray}
\Delta\phi_p=\int_0^T\Omega dt=\int_0^{2\pi}\frac{\Omega}{\dot\phi} d\phi=\langle\theta^2\rangle\frac{\pi k m^2 (4+e^2)}{8a^3(1-e^2)^3}-\langle\eta^2\rangle \frac{\pi a^3\sqrt{1-e^2}}{2m^2k},\label{dp}
\end{eqnarray}
here $a$ is the semi-major axis, $\phi$ is the polar angle. Calculating (\ref{dp}) we use that in the ordinary space $L=mx^2\dot\phi$, $x=a(1-e^2)/(1+e\cos\phi)$, and $p^2/2m-mk/x=-mk/2a$.

Note that in the consequence of violation of the equivalence principle in noncommutative phase space (\ref{for101})-(\ref{for10001}) the perihelion shift depends on the mass of the particle $m$.
It is important to mention that if conditions (\ref{condt}), (\ref{conde}) are satisfied, taking into account  (\ref{thetar2}), (\ref{etar2}) one has
\begin{eqnarray}
{\langle\theta^2\rangle}{m^2}=\frac{3\tilde{\gamma}^2}{2l_P^2}=A,\ \
\frac{\langle\eta^2\rangle}{m^2}=\frac{3\tilde{\alpha}^2}{2l_P^2}=B,\label{cq2}
\end{eqnarray}
where constants $A$, $B$ are the same for particles with different masses. So, substituting  (\ref{cq2}) into expression  (\ref{dp}) one obtains that the perihelion shift does not depend on the particle mass
\begin{eqnarray}
\Delta\phi_p=A\frac{\pi k (4+e^2)}{8a^3(1-e^2)^3}-B \frac{\pi a^3\sqrt{1-e^2}}{2k}.
\end{eqnarray}
Conditions (\ref{condt}), (\ref{conde}) are important for recovering of the weak equivalence principle in noncommutative phase space (\ref{for101})-(\ref{for10001}). The detailed studies of this principle in the frame of rotationally-invariant noncommutative algebra can be found in \cite{GnatenkoEPL18}.

Let us generalize the obtained result for the case of motion of a composite system (macroscopic body) in the gravitational field.
One has the following Hamiltonian
\begin{eqnarray}
H_s=H_{cm}+H_{rel},\\
H_{cm}=\frac{(P^c)^2}{2M}-\frac{Mk}{X^{c}},
 \end{eqnarray}
here $M$ is the total mass of the system, $X^c_i$, $P^c_i$ are coordinates and momenta of the center-of-mass, Hamiltonian $H_{rel}$ corresponds to the relative motion and depends on the relative coordinates.
 On the conditions (\ref{condt}), (\ref{conde}) coordinates and momenta of the center-of-mass satisfy relations of noncommutative algebra (\ref{xc}), (\ref{pc})
and can be represented as $X^c_i=x^c_i-{\theta}^c_{ij}p^c_j/2$, $P^c_i=p^c_i+{\eta}^c_{ij}x^c_j/2$,
where ${\theta}^c_{ij}$, ${\eta}^c_{ij}$ are defined as (\ref{effc}), (\ref{eff2c}).
So, similarly as was considered above for  a particle in the gravitational filed in the case of composite system's motion up to the second order in the parameters of noncommutativity one can study the following Hamiltonian
\begin{eqnarray}
H_0=\langle H_s\rangle_{ab}+H^a_{osc}+H^b_{osc}=\frac{(p^c)^2}{2M}-\frac{Mk}{x^c}+\frac{\langle(\eta^c)^2\rangle(x^c)^2}{12M}-\nonumber\\-\frac{\langle(\theta^c)^2\rangle M k(L^c)^2}{8 (x^c)^5}+\frac{\langle(\theta^c)^2\rangle M k}{24}\left(\frac{1}{(x^c)^2}(p^c)^2\frac{1}{x^c}+\frac{1}{x^c}(p^c)^2\frac{1}{(x^c)^2}+\frac{\hbar^2}{(x^c)^5}\right)+\nonumber\\+\langle H_{rel}\rangle_{ab}+H^a_{osc}+H^b_{osc},\label{h00}
\end{eqnarray}
where ${\bf L}^c=[{\bf x}^c\times{\bf p}^c]$. Taking into account definitions of relative coordinates and relative momenta $\Delta {\bf  X}^{(n)}={\bf X}^{(n)}-{\bf X}^c$, $\Delta{\bf P}^{(n)}={\bf P}^{(n)}-\mu_{n}{\bf P}^c$,  and using  (\ref{condt}), (\ref{conde}), (\ref{rep})  one can write $\Delta X^{(n)}_i=\Delta x^{(n)}_i-\theta^{(n)}_{ij}\Delta p^{(n)}_j/2$, $\Delta P^{(n)}_i=\Delta p^{(n)}_i+\eta^{(n)}_{ij}\Delta x^{(n)}_j/2$ where  $\Delta{\bf  x}^{(n)}={\bf x}^{(n)}-{\bf x}^c$, $\Delta{\bf p}^{(n)}={\bf p}^{(n)}-\mu_{n}{\bf p}^c$  satisfy the ordinary commutation relations. Note that $\langle H_{rel}\rangle_{ab}$ depends on coordinates $\Delta{\bf  x}^{(n)}$ and momenta $\Delta{\bf p}^{(n)}$ and commutes with $H_0$. So, for studies of the classical motion of the center-of-mass of macroscopic body in gravitational field in the noncommutative phase space one can consider the following Hamiltonian
\begin{eqnarray}
\langle H_{cm}\rangle_{ab}=\frac{(p^c)^2}{2M}-\frac{M k}{x^c}+\frac{\langle(\eta^c)^2\rangle(x^c)^2}{12M}-\frac{\langle(\theta^c)^2\rangle M k(L^c)^2}{8 (x^c)^5}+\frac{\langle(\theta^c)^2\rangle M k (p^c)^2}{12(x^c)^3}.
\end{eqnarray}
On the basis of results obtained for a particle (\ref{dp}), the perihelion shift of orbit of macroscopic body in noncommutative phase space reads
\begin{eqnarray}
\Delta\phi_{nc}=\langle(\theta^c)^2\rangle\frac{\pi k M^2 (4+e^2)}{8a^3(1-e^2)^3}-\langle(\eta^c)^2\rangle \frac{\pi a^3\sqrt{1-e^2}}{2M^2k},\label{df}
\end{eqnarray}
where
\begin{eqnarray}
{\langle(\theta^{c})^2\rangle}=\frac{3\tilde{\gamma}^2}{2l_P^2M^2}=\frac{A}{M^2},\label{cq1n}\\
{\langle(\eta^c)^2\rangle}=\frac{3\tilde{\alpha}^2M^2}{2l_P^2}=B M^2.\label{cq2n}
\end{eqnarray}

\section{Estimation of the upper bounds on the parameters of noncommutativity}\label{rozd3}
 To estimate the upper bounds for the parameters of noncommutativity let us apply the obtained result for the Mercury planet. Observed precession of perihelion of the Mercury planet which cannot be explained by the Newtonian gravitational effects of other planets and asteroids, Solar Oblateness is \cite{Park}
\begin{eqnarray}
\Delta\phi_{obs}=42.9779\pm0.0009\textrm{ arc-seconds per century}=\nonumber\\=2\pi(7.98695\pm0.00017)\cdot10^{-8}\textrm{radians/revolution}
 \end{eqnarray}
 This advance is explained by relativistic effects such as Lense-Thirring and gravitoelectric effect \cite{Park}.
Similar as was done in \cite{Benczik02} for estimation of the minimal length in the deformed space, and in \cite{Romero,Mirza,Djemai2,Gnatenko_arxiv} for estimation of the minimal length in noncommutative space of canonical type we compare the perihelion shift caused by noncommutativity (\ref{df}) with $\Delta\phi_{obs}-\Delta\phi_{GR}=2\pi (-0.00049\pm 0.00017)\cdot 10^{-8}\textrm{radians/revolution}$ (here
$\Delta\phi_{GR}=2\pi(7.98744 \cdot 10^{-8})$radians/revolution  is the perihelion precession rate from General Relativity predictions). Assuming that $|\Delta\phi_{nc}|$ is less than $|\Delta\phi_{obs}-\Delta\phi_{GR}|$ at $3\sigma$ we have
\begin{eqnarray}
|\Delta\phi_{nc}|\leq 2\pi\cdot 10^{-11}\textrm{radians/revolution},\label{eq}
\end{eqnarray}
where $\Delta\phi_{nc}$ is given by (\ref{df}) with  $k=GM_{\odot}$ ($G$ is the gravitational constant, $M_{\odot}$ is the mass of the Sun).
Since $\theta_{ij}^{c}$ or $\eta_{ij}^{c}$ could be equal to zero, to estimate the orders of parameters of noncommutativity it is sufficiently to consider the following inequalities
\begin{eqnarray}
\left|\langle(\theta^c)^2\rangle\frac{\pi G M_{\odot} M^2 (4+e^2)}{8a^3(1-e^2)^3}\right|\leq 2\pi\cdot10^{-11}\textrm{radians/revolution},\\
\left|\langle(\eta^c)^2\rangle \frac{\pi a^3\sqrt{1-e^2}}{2G M_{\odot}M^2}\right|\leq 2\pi\cdot10^{-11}\textrm{radians/revolution},
\end{eqnarray}
(here $M$ is the mass of Mercury, $a$, $e$ are parameters of its orbit) from which one finds
\begin{eqnarray}
\hbar\sqrt{\langle(\theta^{c})^2\rangle}<2.3\cdot10^{-57}\textrm{m}^2,\label{ucm1n}\\
\hbar\sqrt{\langle(\eta^{c})^2\rangle}<1.8\cdot10^{-22}\textrm{kg}^2\textrm{m}^2/\textrm{s}^2.\label{ucm2n}
\end{eqnarray}

 On the basis of relations (\ref{cq2}), (\ref{cq1n}), (\ref{cq2n}), we can write
\begin{eqnarray}
\langle(\theta^{c})^2\rangle M^2=\langle(\theta^{(n)})^2\rangle m_n^2, \label{rr1}\\
\frac{\langle(\eta^{c})^2\rangle}{M^2}=\frac{\langle(\theta^{(n)})^2\rangle}{m_n^2},\label{rr2}
\end{eqnarray}
where $\langle(\theta^{(n)})^2\rangle$, $\langle(\eta^{(n)})^2\rangle$ are given by (\ref{thetar2}), (\ref{etar2}) and correspond to particle of mass $m_n$.
 Note that relations (\ref{rr1}), (\ref{rr2}) with $m_n$, $M$ being measured masses follows from the weak equivalence principle \cite{GnatenkoEPL18}.

So, using (\ref{ucm1n}), (\ref{ucm2n}) and taking into account relations (\ref{rr1}), (\ref{rr2}) we can estimate  upper bounds on the parameters of noncommutativity corresponding to particles. Upper bounds for the parameters of noncommutativity corresponding to the electron read $\hbar\sqrt{\langle(\theta^{(e)})^2\rangle}<8.3\cdot10^{-4}\textrm{m}^2$ and
\begin{eqnarray}
\hbar\sqrt{\langle(\eta^{(e)})^2\rangle}<5.1\cdot10^{-76}\textrm{kg}^2\textrm{m}^2/\textrm{s}^2.\label{R9}
\end{eqnarray}
The obtained upper bound for the parameter of coordinate noncommutativity is not strong. This is because of reduction of value $\langle(\theta^{c})^2\rangle$ with respect to $\langle(\theta^{(n)})^2\rangle$ corresponding to individual particle. Namely, from (\ref{rr1}) we have $\langle(\theta^{c})^2\rangle=\langle(\theta^{(n)})^2\rangle m_n^2/M^2$. Note, that in the case when composite system of mass $M$ is made of $N$ particles with the same masses $m$ we can write $\langle(\theta^{c})^2\rangle=\langle\theta^2\rangle/N^2$, where $\langle\theta^2\rangle$ corresponds to a particle.  Therefore influence of noncommutativity of coordinates on the motion of macroscopic bodies is less than this influence on the motion of particles. So, to find stringent upper bound on the parameters of coordinate noncommutativity, studying motion of macroscopic bodies in noncommutative phase space,  experimental data with very hight accuracy are needed.

Upper bound on the parameter of momentum noncommutativity (\ref{R9}) is stringent. It is at least ten orders less than that obtained on the basis of studies of the hydrogen atom in noncommutative phase space without preserved rotational symmetry \cite{Bertolamih}, and on the basis of studies of the hydrogen and exotic atoms in rotationally-invariant noncommutative phase space \cite{GnatenkoIJMPA17,GnatenkoIJMPA18}.
Taking into account (\ref{R9}), for the minimal momentum we can write
\begin{eqnarray}
p_{min}=\sqrt[4]{\frac{3\hbar^2\langle(\eta^{(e)})^2\rangle}{2}}<2.5\cdot10^{-38}\textrm{kg}\cdot\textrm{m}/\textrm{s}.\label{mm}
\end{eqnarray}
In (\ref{mm}) we use expression for the minimal momentum in the rotationally-invariant noncommutative phase space obtained in \cite{Shyiko}.

Similarly, on the basis of result (\ref{ucm2n}) we can estimate upper bound for the parameter of momentum noncommutativity corresponding to nucleons.
 Taking into account that ${\langle(\eta^{c})^2\rangle}/{M^2}={\langle(\theta^{(nuc)})^2\rangle}/{m_{nuc}^2}$ (here $m_{nuc}$ is the mass of nucleon) we find
\begin{eqnarray}
\hbar\sqrt{\langle(\eta^{(nuc)})^2\rangle}<9.3\cdot10^{-73}\textrm{kg}^2\textrm{m}^2/\textrm{s}^2.\label{R99}
\end{eqnarray}
This result  is not so stringent as was obtained in the case of studies of perihelion shift of the Mercury planet in noncommutative phase space of canonical type (upper bound (\ref{R99}) is 7 orders greater than presented in \cite{Gnatenko_arxiv}). This is because in contrast to noncommutative phase space of canonical type expression for perihelion shift of the Mercury planet (\ref{df})  does not contain terms of the first order in the parameters of noncommutativity (these terms vanish after averaging over the eigenfunctions of the harmonic oscillators, see (\ref{he})).  We would like also to note, that upper bound (\ref{R99}) is $6$ orders less than that
obtained on the basis of studies of neutrons in gravitational quantum well in noncommutative phase space of canonical type \cite{Bertolami05}.

\section{Conclusions}

We have considered noncommutative phase space of canonical type with preserved rotational and time reversal symmetries which is characterized by relations (\ref{for101})-(\ref{for10001}).

Influence of noncommutativity of coordinates and noncommutativity of momenta on the perihelion shift of orbit of a particle in Coulomb potential in the space is found up to the second order in the parameters of noncommutativity.
Taking into account features of description of motion of macroscopic body in noncommutative phase space (\ref{for101})-(\ref{for10001}), the result has been applied for the case of the Mercury planet.  Comparing the obtained expression for the perihelion shift of the Mercury planet caused by noncommutativity (\ref{df}) with data for precession of Mercury's perihelion from ranging to the MESSENGER spacecraft the upper bounds for the effective parameters of coordinate and momentum noncommutativity have been estimated (\ref{ucm1n}), (\ref{ucm2n}). The results have been reexamined for the parameters of noncommutativity corresponding to electrons and nucleons.

 We have obtained quite stringent  upper bounds for the parameters of momentum noncommutativity (\ref{R9}), (\ref{R99}) and for the minimal momentum (\ref{mm}).   The result for the parameter of momentum noncommutativity corresponding to the electrons (\ref{R9}) is at least 10 orders less than that obtained on the basis of studies of the hydrogen atom  in the rotationally-invariant noncommutative phase space \cite{GnatenkoIJMPA17} and in noncommutative phase space without preserved rotational symmetry \cite{Bertolamih}.

\section*{Acknowledgments}
The author thanks Prof. Tkachuk V. M. for his advices and support during research studies.   Publication contains the results of studies conducted by President's of Ukraine grant for competitive projects ($\Phi$-82). This work was partly supported by the Project $\Phi\Phi$-63Hp (No. 0117U007190) from the Ministry of Education and Science of Ukraine.

\end{document}